\documentclass[12pt]{article}

\usepackage{a4wide,graphics,graphicx,amsmath,amssymb,cite,nicefrac}
\usepackage[verbose]{wrapfig}
\usepackage{authblk,hyperref}
\catcode`@=11 \@addtoreset{equation}{section} \catcode`@=12

\begin{document}
\date{}
\title{
{\baselineskip -.2in
\vbox{\small\hskip 4in \hbox{IITM/PH/TH/2014/2}}
} 
\vskip .4in
\vbox{
{\bf \LARGE Multiple Single-Centered Attractors}
}}
\author[1]{Pramod Dominic\thanks{email:pramod@uccollege.edu.in}}
\author[2]{Taniya Mandal\thanks{email: taniya@physics.iitm.ac.in}}
\author[2]{Prasanta K. Tripathy\thanks{email: prasanta@physics.iitm.ac.in}}

\affil[1]{\normalsize\it Department of Physics, \authorcr \it Union Christian College, \authorcr \it Aluva, Kerala 683102, India.}
\affil[2]{\normalsize\it Department of Physics, \authorcr \it Indian Institute of Technology Madras, \authorcr \it Chennai 600036, India.}
\maketitle
\begin{abstract}

In this paper we study spherically symmetric single-centered attractors in $\mathcal N=2$ supergravity in four 
dimensions. The attractor points are obtained by extremising  the effective black hole potential in the moduli 
space. Using the 4D-5D correspondence of critical points of the effective black hole potential we argue the 
existence of multiple attractors in four dimensions and explicitly construct a pair of multiple solutions in a simple 
two parameter model. We further obtain explicit examples of two distinct non-supersymmetric attractors  in type 
$IIA$ string theory compactified  on $K3\times T^2$ carrying $D0-D4-D6$ charges. We compute the entropy 
of these attractors and analyse their stability  in detail.

\end{abstract}

\newpage

\section{Introduction}

The attractor mechanism, discovered originally by the seminal work of Ferrara, Kallosh and 
Strominger \cite{Ferrara:1995ih} in studying supersymmetry preserving, static, spherically 
symmetric, magnetically charged black holes in four-dimensional
$\mathcal N=2$ supergravity theory coupled to vector multiplets,  plays a central role in understanding 
the origin of black hole entropy. For these supersymmetric black holes, as one approaches the
 horizon, the scalar fields run into a fixed point in the moduli space. The fixed point in the 
moduli space is solely determined by the black hole charges and is independent of the values 
of the scalar fields at spatial infinity. Though initially it used to be thought as a consequence of 
supersymmetry, it has been subsequently shown that, for the single-centered, static, spherically 
symmetric configurations, the attractor mechanism is a consequence of extremality of the black 
hole \cite{Ferrara:1997tw}. This gives rise to the possibility of exploring the properties of extremal 
black holes which are non-supersymmetric \cite{Goldstein:2005hq}. 

Though there has been extensive study of non-supersymmetric attractors in recent years
(see Refs.\cite{Ferrara:2008hwa,Bellucci:2007ds} for a review on the topic), there are still a 
number of issues which remain to be resolved. One such issue is the stability of such black 
holes. Unlike their supersymmetric counterparts which are guaranteed to be stable, these
non-supersymmetric attractors can either be stable or unstable \cite{Tripathy:2005qp,Nampuri:2007gv}. 
For a large class of models, including the ones arising from string compactifications, they also 
possess flat directions \cite{Tripathy:2005qp}. 
A general group theoretic analysis has been carried out in \cite{Ferrara:2007pc,Ferrara:2007tu}
to understand the issue of stability 
and the existence of flat directions for non-supersymmetric attractors in more general class of $\mathcal N=2$
supergravity theories.
It has been subsequently shown that, stringy corrections can make these flat directions 
either stable or unstable depending upon the charges of the corresponding black hole configurations
\cite{Bellucci:2007eh,Bellucci:2008tx,Dominic:2011st,Bellucci:2010zd,Dominic:2011iz,Dominic:2010yv}.

One of the important issues with regard to the non-supersymmetric attractors is the construction of 
a fake superpotential for them. A large class of extremal non-supersymmetric attractors can be 
obtained upon extremising a suitably constructed fake superpotential \cite{Ceresole:2007wx,
Dall'Agata:2011nh,Andrianopoli:2007gt,Ceresole:2010nm,Bossard:2009we}. Interestingly all 
such non-supersymmetric attractors are axion free and are related by a ${\mathbb Z}_2$ symmetry
to their respective supersymmetric cousins. Construction of more general non-supersymmetric 
attractors with non-vanishing axionic part from suitable fake superpotentials remains to be explored.

One other related issue of interest is the multiplicity of these attractors \cite{Moore:1998zu,Moore:1998pn}. 
It is well known that, for extremal black holes, the equations of motion become algebraic as one 
approaches the  horizon. For the static, spherically symmetric case, the black hole is described in terms 
of the motion of a particle in an effective one dimensional theory. The attractor value in the moduli space 
is obtained by extremising the effective black hole potential, which is an algebraic function of the vector 
multiplet moduli \cite{Goldstein:2005hq,Kallosh:2005ax,Kallosh:2006bt,Ceresole:2007rq}.  It is indeed 
possible to obtain single centered multiple attractors by solving these algebraic equations. Supersymmetric 
multiple attractors have already been constructed in a  simple two-parameter model consisting of the 
five dimensional $\mathcal N=2$ ungauged supergravity coupled to two abelian vector multiplets 
\cite{Kallosh:1999mz}. Multiple solutions in the context of flux compactification and their connection with the 
`area code' for $\mathcal N=2$ supergravity with non-homogeneous scalar manifold has been explored in 
\cite{Giryavets:2005nf,Misra:2007yu}.

Though there has been some progress on five dimensional supersymmetric multiple attractors, not 
much work has been carried out in studying single centered multiple supersymmetric as well as  
non-supersymmetric attractors in four dimensions. The focus of our current investigation is to study 
these four dimensional multiple attractors. Using the $4D-5D$ correspondence of black hole attractor 
points \cite{Ceresole:2007rq} we  construct multiple attractors in four dimensional $\mathcal N=2$ supergravity 
theory arising from the compactification of type $IIA$ supergravity on a Calabi-Yau manifold.  We also 
discuss multiple axionic non-supersymmetric attractors which have no obvious five dimensional origin.

The plan of the paper is as follows. In the next section we will review the basics of attractor mechanism
in four dimensional ${\cal N}=2$ theories. Subsequently, in \S3 we will obtain four dimensional axion free 
multiple black hole configurations using the $4D-5D$ correspondence.  We will then focus on multiple
non-supersymmetric attractors with non-vanishing axionic part in \S4. Here we will construct the attractors
by explicitly solving the equations of motion and discuss their stability. Finally in \S5 we will summarise 
our findings. Some of the detail calculations will be carried out in the appendices.

\section{Background}

In this section we will review the basics of attractor mechanism. We consider ${\cal N}=2$ supergravity 
theory in four dimensions coupled to $n$ vector multiplets. Hypermultiplets do not play any role in our
analysis. The bosonic part of the  supergravity Lagrangian is given by:
\begin{eqnarray}\label{sugra}
{\cal L} = - \frac{R}{2} +  g_{a\bar b} \partial_\mu x^a \partial_\nu \bar x^{\bar b} h^{\mu\nu} 
-  \mu_{\Lambda\Sigma} {\cal F}^\Lambda_{\mu\nu}{\cal F}^\Sigma_{\lambda\rho} 
h^{\mu\lambda}h^{\nu\rho} -   \nu_{\Lambda\Sigma} {\cal F}^\Lambda_{\mu\nu}
*{\cal F}^\Sigma_{\lambda\rho} h^{\mu\lambda}h^{\nu\rho} 
\end{eqnarray}
Our notations and conventions in this section are same as in Ref\cite{Ferrara:1997tw}. The vector 
moduli are denoted by the $n$ complex scalars $x^a$ with moduli space metric $g_{a\bar b}$.
The vector fields $A^\Lambda, (\Lambda = 0,1,\cdots,n)$ with the corresponding field strengths 
${\cal F}^\Lambda$ consists of the graviphoton as well as  the gauge fields from the vector multiplets. 
$\mu_{\Lambda\Sigma}$ and $\nu_{\Lambda\Sigma}$ are the gauge couplings, $h_{\mu\nu}$ is 
the metric of the four dimensional space time with scalar curvature $R$. The moduli space metric 
as well as the gauge couplings are determined in terms of the ${\cal N}=2$ prepotential $F$.

For static, spherically symmetric configurations the metric $h_{\mu\nu}$ is given by:
\begin{eqnarray}
ds^2 = e^{2U} dt^2 - e^{-2U}\gamma_{mn} dy^mdy^n
\end{eqnarray}
where, for the extremal black holes, the spatial part of the above is given by the Euclidean metric 
$\gamma_{mn} =  \delta_{mn}$, and the warp factor $U$ depends only on the radial coordinate $r$. 
Substituting  this ansatz, and an appropriate expression for the gauge field $A^\Lambda$ satisfying 
the Bianchi identity, in the field equations for the Lagrangian (\ref{sugra}), we find that they are
equivalent to the equations of motion of  an effective one dimensional system whose Hamiltonian
is constrained to be zero. For regular black hole horizon, the effective potential of the one dimensional
system must be extremized at the black hole horizon.  Since the effective black hole potential is a 
function of the scalar fields as well as the black hole charges, generically the extremization fixes 
the scalar fields at the horizon and their values are determined by the black hole charges. For 
stable attractors the Hessian must admit positive eigenvalues. 

Our focus in this paper is on ${\cal N}=2$ supergravity theories arising from the compactification of 
type $IIA$ string theory on a Calabi-Yau manifold ${\cal{M}}$. In the present work we will throughout 
work using the special coordinate basis which admits a holomorphic prepotential.
{\footnote{This is not the case always and there exists string compactifications without admitting 
a special coordinate basis as has been first shown in \cite{Ceresole:1995jg}.}}
 In the large volume limit, the  ${\cal N} = 2$ prepotential is given by
 \begin{equation} \label{prepot}
F=D_{abc} \frac{X^a X^b X^c}{X^0}
\end{equation}
where $D_{abc}$ are the intersection numbers: 
$D_{abc} = (\nicefrac{1}{6})\int_{\cal M} \alpha_a\wedge\alpha_b\wedge\alpha_c$ with 
$\alpha_a\in H^2({\cal M},\mathbb{Z})$. We consider dyonic charged black holes with 
electromagnetic charges $(p^\Lambda,q_\Sigma)$ arising due to $D$-branes wrapped on various
cycles of the Calabi-Yau manifold. The effective black hole potential in this case can be expressed
as \cite{Ferrara:1997tw}:
\begin{equation}
V=e^K \Big[g^{a \bar{b}} \nabla_a W \overline{\nabla_b W}+|W|^2\Big] \ . 
\end{equation}
%
 %
Here, $K$ denotes the K\"{a}hler potential and $W$ is the superpotential. The K\"ahler covariant
derivative is defined as $\nabla_a W=\partial_a W+\partial_a K W$ and the moduli space metric
is given by $g_{a\bar b} = \partial_a\partial_{\bar b} K$. The superpotential $W$ is determined 
by the prepotential $F$ and the dyonic charges of the black hole:
\begin{eqnarray}
W = \sum_{\Lambda=0}^n (q_\Lambda X^\Lambda-p^\Lambda \partial_\Lambda F) \ , 
\end{eqnarray}
where as the K\"ahler potential is given by:
\begin{equation}
K =  -\log\Big[i\sum_{\Lambda=0}^{n} (\overline{X^\Lambda} \partial_\Lambda F 
- X^\Lambda \overline{\partial_\Lambda F})\Big] \ . 
\end{equation}

The attractor values are determined by extremising the effective black hole potential \cite{Bellucci:2006ew,Bellucci:2006xz}.
Upon setting $\partial_a V = 0$, we obtain:
\begin{eqnarray} \label{nonsusy}
g^{b\bar{c}} \nabla_a\nabla_b W \overline{\nabla_c W}
+2 \nabla_a W \overline{W}+\partial_a g^{b\overline{c}}\nabla_b W\overline{\nabla_c W} = 0 \ .
\end{eqnarray}
This is the necessary condition to have a regular horizon. This equation admits both supersymmetric as well
as non-supersymmetric solutions. Supersymmetric solutions satisfy $\nabla_a W = 0$, where as the 
solutions for which $\nabla_a W\neq 0$ give rise to non-supersymmetric  attractors.

The supersymmetry preserving attractors are guaranteed to be stable\cite{Ferrara:1997tw}. However for the non-supersymmetric attractors 
this is not the case in general. In the later case the stability is ensured only when the mass matrix corresponding to the 
effective black hole potential admits non-negative eigenvalues \cite{Goldstein:2005hq,Tripathy:2005qp}. 

\section{Multiple Axion-free Attractors}

Multiple basin of attractors have been first constructed in the context of ungauged $\mathcal N=2$ supergravity
theory in five dimensions  \cite{Kallosh:1999mz}. In the simplest case of two vector multiplet moduli,
the authors constructed multiple supersymmetric attractor points with the same charge configurations 
which are related to one another by a $\mathbb{Z}_2$ symmetry. In this section, we will first review these 
attractor solutions. It is indeed possible to construct multiple attractors in four dimensions from these
solutions, by using the well known $4D$-$5D$ correspondence of black hole critical points 
\cite{Ceresole:2007rq}.  Some of these solutions are extensively studied in literature where as the others 
lead to genuinely new solutions in four dimensions. 

We will consider five dimensional ungauged $\mathcal N=2$ supergravity coupled to abelian vector multiplets
\cite{Gunaydin:1983bi,Gunaydin:1984ak}. This supergravity, for example, may be obtained upon the
compactification of the M-theory on a Calabi-Yau manifold. The real scalar fields $\hat\lambda^a$
span the vector multiplet moduli space. In five dimensional supergravity they are constrained by the
relation:
\begin{equation}\label{constrnt}
 D_{abc} \hat\lambda^a\hat\lambda^b\hat\lambda^c = 1 \ , 
\end{equation}
where, as usual, $D_{abc}$ denote the triple intersection numbers of a basis of two cycles in 
$H_2({\cal M},\mathbb Z)$ of  the Calabi-Yau manifold ${\cal M}$. Here we will mainly  focus on the 
examples of the multiple attractors in five dimensions constructed in \cite{Kallosh:1999mz} and unless 
otherwise specified, in this section we will closely follow their notations and conventions.

The supersymmetric attractors in five dimensions with charge configuration $\{\hat q_a\}$ are obtained 
by extremizing the corresponding $\mathcal N=2$ central charge:
\begin{equation}\label{fivedcc}
Z_5 = \hat q_a \hat\lambda^a \ . 
\end{equation}
Extremising the central charge (\ref{fivedcc}), with respect to the scalar fields $\hat\lambda^a$
subjected to the constraint (\ref{constrnt}), we find
\begin{equation}\label{fivedccx}
 \hat q_a-D_{abc}\hat{\lambda}^b\hat{\lambda}^c(\hat{\lambda}^d \hat q_d)=0 \  .
\end{equation}
One can introduce variables $\bar\lambda^a = \hat\lambda^a \sqrt{\hat q_b\hat\lambda^b}$, in terms of which
the above equation can be written as 
\begin{equation}\label{fivedcxx}
 \hat q_a-D_{abc}\bar{\lambda}^b\bar{\lambda}^c =0 \  .
\end{equation}
Thus we obtain a set of quadratic coupled equations in terms of the variables $\bar\lambda^a$.
Solutions to the equations of motion (\ref{fivedcxx}) have been studied extensively in \cite{Kallosh:1999mz}.
For the simplest case of a two parameter model, we have a  system of coupled quadratic equations in two 
variables. The general solution to this system can be more compactly expressed upon setting $D_{111}={a}$, 
$D_{112}={b}$, $D_{122}={c}$ and $D_{222}={d}$ and by introducing the notations 
$\mathcal{L}=ad - bc$, $\mathcal{M}={c}^2- bd $, $\mathcal{N}= {b}^2- ac$, 
$\mathcal{D}=\mathcal{M}q_1^2+\mathcal{N} q_2^2+\mathcal{L} q_1 q_2$,
$\mathcal{E}= {c} q_1- {b} q_2$, $\mathcal{F}={d} q_1- {c} q_2$ and $\mathcal{H}= {b} q_1- {a} q_2$.
We find \cite{Kallosh:1999mz}:
\begin{eqnarray}\label{multsol5d}
(\bar{\lambda}^1_{\pm})^2 &=& \frac{\mathcal{F} \mathcal{L}
+2 \mathcal{E} \mathcal{M}}{\mathcal{L}^2
-4 \mathcal{M} \mathcal{N}} \pm \frac{\sqrt{4 \mathcal{M}^2 \mathcal{D}}}{\mathcal{L}^2-4 \mathcal{M} \mathcal{N}},\cr
(\bar{\lambda}^2_{\pm})^2 &=& -\frac{\mathcal{H} \mathcal{L}
+2 \mathcal{E} \mathcal{N}}{\mathcal{L}^2
-4 \mathcal{M} \mathcal{N}} \pm \frac{\sqrt{4 \mathcal{N}^2 \mathcal{D}}}{\mathcal{L}^2-4 \mathcal{M} \mathcal{N}} .
\end{eqnarray}
Clearly, the pair of solutions are related one another by a $\mathbb{Z}_2$ symmetry. As explained in detail in 
\cite{Kallosh:1999mz}, the right hand side is positive definite for one of the solutions (the one corresponding 
to negative sing of the square root), where as it is negative definite for the other one. Hence, the values of 
$\bar\lambda^i$ for the corresponding solution $(\bar\lambda_+^1,\bar\lambda_+^2)$ are imaginary. However,
this is not an issue because the physics fields corresponding to both the solutions 
 $(\hat\lambda_\pm^1,\hat\lambda_\pm^2)$ are always real.

There exists a natural correspondence between the attractor points in five dimensions with the  the critical points 
of the black hole effective potential for axion free configurations in four dimensions  \cite{Ceresole:2007rq}. Using 
this $4D$-$5D$ correspondence the five dimensional attractor equation can be mapped to the corresponding 
equation for the four dimensional axion free black hole attractors. In the following we will analyse multiple axion
free attractors in four dimensions using this correspondence.

To demonstrate this, let us 
consider, for example, the $D2-D6$ configuration in the compactification of  type $IIA$ supergravity
on a Calabi-Yau manifold. Using the tree-level prepotential (\ref{prepot}) and the expressions for the 
K\"ahler potential $K$ and superpotential $W$ as defined in the
previous section, we find:
\begin{eqnarray} \label{khlr}
K &=& -\log\Big[- i D_{abc}(x^a-\bar{x}^a) (x^b-\bar{x}^b) (x^c-\bar{x}^c)\Big],\\
W &=& q_a x^a+ p^0 D_{abc}x^ax^bx^c.
\end{eqnarray}
We have denoted the $D2$ charges by $q_a$ and the $D6$ charge by $p^0$.
For convenience, we have introduced here the notation  $x^a = X^a/X^0$ and, exploiting the 
symplectic invariance, we set the gauge $X^0=1$. The supersymmetric attractors satisfy $\nabla_a W =0$. 
For the $D2-D6$ configuration, this condition
can straightforwardly be evaluated. For the axion free configurations, we find
\begin{equation} \label{d2d6a}
q_a-\frac{3p^0}{2}D_{abc}\lambda^b\lambda^c-\frac{3}{2\nu}D_{abc}\lambda^b\lambda^c\lambda^{d} q_d=0,
\end{equation}
where the real scalar field $\lambda^a$ is the axion free part of $x^a$ (i.e. $x^a = i \lambda^a$). In
addition, we use the notation $\nu= D_{abc}\lambda^a\lambda^b\lambda^c$. Multiplying $\lambda^a$ and 
summing over we find 
$$ q_a\lambda^a + 3 p^0 \nu = 0 \ . $$
Using the above Eq.({\ref{d2d6a}) can be simplified to obtain a set of coupled quadratic equations:
\begin{equation} \label{d2d6}
q_a + 3 p^0 D_{abc}\lambda^b\lambda^c = 0  \ .
\end{equation}
This equation reduces to Eq.(\ref{fivedcxx}) upon the identification $\hat q_a = q_a$ and 
$\bar\lambda^a = \sqrt{-3 p^0} \lambda^a$ \cite{Ceresole:2007rq}. However, this does not imply the existence of 
a corresponding solution to the four dimensional supersymmetric  equations (\ref{d2d6}) for every solution of the five 
dimensional supersymmetric conditions (\ref{fivedcxx}). The nonlinear relation $\bar\lambda^a = \hat\lambda^a 
\sqrt{\hat q_b\hat\lambda^b}$ allows the possibility of having complex values for $\bar\lambda^a$ as long as 
$\hat\lambda^a$ is real where as the solutions of Eq.(\ref{d2d6}) for the four dimensional fields $\lambda^a$ 
must be real. 

This does not rule out the possibility of admitting multiple four dimensional attractors for a given set of charges
because of the existence of non-supersymmetric attractors. For the $D2-D6$ configuration, the non-supersymmetric 
attractor can be constructed from a fake superpotential: 
\begin{equation}
W = q_a x^a -  p^0 D_{abc}x^ax^bx^c.
\end{equation}
Upon extermisation this gives 
\begin{equation}
q_a+\frac{3p^0}{2}D_{abc}\lambda^b\lambda^c-\frac{3}{2\nu}D_{abc}\lambda^b\lambda^c\lambda^{d} q_d=0,
\end{equation}
As before, this equation can be rewritten as 
\begin{equation} \label{d2d6ns}
q_a - 3 p^0 D_{abc}\lambda^b\lambda^c = 0  \ .
\end{equation}
The above equation corresponds to five dimensional supersymmetric attractor upon the identification 
$\hat q_a = q_a$ and $\bar\lambda^a = \sqrt{3 p^0} \lambda^a$. Thus, for a given set of $\hat q_a$ if there 
exists two physical solutions of the five dimensional equations with  one of them admitting real values
for $\bar\lambda^a$  where as the other one having pure imaginary $\bar\lambda^a$, then one of these 
two solutions will correspond to a supersymmetric attractor where as the other one will be non-supersymmetric
 in four dimensions. Both the solutions will exist in four dimensions for a given value of $D2-D6$ charges.
 This is contrary to the previously known examples where the supersymmetric and non-supersymmetric attractors
 existed in mutually exclusive domains in the charge lattice \cite{Tripathy:2005qp,Ceresole:2007rq}. 

This can be best understood in the case of two parameter model. Using the five dimensional solutions 
(\ref{multsol5d}) the exact analytic expression for the solutions to both the supersymmetric as well as 
non-supersymmetric equations of motion can be written down explicitly.
For the supersymmetric equations (\ref{d2d6}), the solutions are given by
\begin{eqnarray}\label{multsolsusy}
({\lambda}^1_{s\pm})^2 &=&\frac{1}{3 p^0(4 {\mathcal{M} \mathcal{N}}-\mathcal{L}^2)}
\Big(  ({\mathcal{F} \mathcal{L} +2 \mathcal{E} \mathcal{M}})
\pm \sqrt{4 \mathcal{M}^2 \mathcal{D}}\Big) , \cr
({\lambda}^2_{s\pm})^2 &=& \frac{1}{3 p^0(4 {\mathcal{M} \mathcal{N}}-\mathcal{L}^2)}
\Big( - (\mathcal{H} \mathcal{L}+2 \mathcal{E} \mathcal{N}) 
\pm \sqrt{4 \mathcal{N}^2 \mathcal{D}} \Big) .
\end{eqnarray}
where as the non-supersymmetric solutions obtained from Eq.(\ref{d2d6ns}) are
\begin{eqnarray}\label{multsolnsusy}
({\lambda}^1_{ns\pm})^2 &=&\frac{1}{3 p^0({\mathcal{L}^2-4 \mathcal{M} \mathcal{N}})}
\Big(  ({\mathcal{F} \mathcal{L} +2 \mathcal{E} \mathcal{M}})
\pm \sqrt{4 \mathcal{M}^2 \mathcal{D}}\Big) , \cr
({\lambda}^2_{ns\pm})^2 &=& \frac{1}{3 p^0({\mathcal{L}^2-4 \mathcal{M} \mathcal{N}})}
\Big( - (\mathcal{H} \mathcal{L}+2 \mathcal{E} \mathcal{N}) 
\pm \sqrt{4 \mathcal{N}^2 \mathcal{D}} \Big) .
\end{eqnarray}
To analyse the above solutions, note that 
\begin{eqnarray*}
({\mathcal{F} \mathcal{L} +2 \mathcal{E} \mathcal{M}})^2 - 4 \mathcal{M}^2 \mathcal{D} 
= \mathcal{F}^2 ({\mathcal{L}^2-4 \mathcal{M} \mathcal{N}}) , \cr
(\mathcal{H} \mathcal{L}+2 \mathcal{E} \mathcal{N})^2 - 4 \mathcal{N}^2 \mathcal{D}
= \mathcal{H}^2 ({\mathcal{L}^2-4 \mathcal{M} \mathcal{N}}) .
\end{eqnarray*}
Consider, for example, the case $(4 {\mathcal{M} \mathcal{N}}-\mathcal{L}^2) >0$. We can see that, in this case,
 for a given set of charges $(q_a,p^0)$, if $(\lambda^1_{s+})^2 >0$, then $(\lambda^2_{s+})^2, 
(\lambda^1_{ns-})^2, (\lambda^2_{ns-})^2$ are all positive and $(\lambda^1_{s-})^2, (\lambda^2_{s-})^2, 
(\lambda^1_{ns+})^2, (\lambda^2_{ns+})^2$ are all negative and vice versa. Thus we have multiple 
attractors in the entire domain of the charge lattice.

Existence of these multiple attractors might appear inconsistent with the uniqueness results of black 
hole attractors \cite{Wijnholt:1999vk}. However, as explained in a subsequent work by Kallosh 
\cite{Kallosh:1999mb}, there is no contradiction since the moduli space has disjoint branches and the
attractors are unique in each  of these branches.

\section{Multiple Non-supersymmetric Attractors}
In this section we will discuss more general solutions of the attractor condition (\ref{nonsusy}) in detail. For 
simplicity, we will consider the case where the non-vanishing intersection numbers  are of the form $D_{abs}$, 
where 
the index $s$ is taking a fixed value $s=n$ and the indices $a,b = 1,2,\cdots,(n-1)$, i.e., 
topologically the Calabi-Yau manifold must be a product form of the type ${\cal M} = {\cal M}_4 \times {\cal M}_2$. 
To be more specific we consider the example of $K3\times T^2$. However, our analysis  will also be valid for 
$T^6$ since the intersection numbers satisfy the above property in this case as well. We consider the 
${\cal N}=2$ truncation of the type $IIA$ string compactification on the above manifolds.

The prepotential (\ref{prepot}) now takes the form:
\begin{equation}\label{prepotx}
F=C_{a b} \frac{X^a X^b X^s}{X^0} \ , 
\end{equation}
where $C_{ab}$ is the intersection matrix for a basis of $H_2(K3,{\mathbb{Z}})$: the integral homology 
of $2$-cycles in $K3$. 
As in the previous section, we introduce the notation $x^a = X^a/X^0, x^s = X^s/X^0$  as well as  set the 
gauge $X^0=1$ now on. We will consider black hole solutions arising due to the intersecting $D0-D4-D6$
configurations. The K\"ahler potential and the superpotential for this configuration are given by 
\begin{eqnarray}
K &=& -\log\Big[-i C_{a b} (x^a-\bar{x}^a) (x^b-\bar{x}^b) (x^s-\bar{x}^s)\Big]\\
W &=& q_0 -2 p^a C_{a b} x^b x^s-p^s C_{a b} x^a x^b +p^0 C_{a b} x^a x^b x^s
\end{eqnarray}
The metric on the moduli space is found to be 
\begin{eqnarray}
g_{s \bar{s}} =  -\frac{1}{(x^s-\bar{x}^s)^2}, \ 
g_{a\bar b} = \frac{2}{M}\left( C_{ab} - \frac{2}{M} M_a M_b\right), \ {\rm and} \ g_{a\bar s} = 0 \ ,
\end{eqnarray}
where we have introduced the notation $M_a = C_{ab}(x^b - \bar x^b)$ and $M = M_{a}(x^a - \bar x^a)$ for 
convenience. We will now outline some of the intermediate steps to carry out the equations of motion. Some 
of the individual terms contributing to the equations of motion are:
\begin{align*}
 \nonumber \nabla_a W &= 2 \big(p^0 C_{ab} x^b x^s-D_a x^s -C_{ab} x^b p^s\big)-\frac{2 M_a}{M}W,\\
 \nonumber \nabla_s W &= p^0 C_{ab} x^a x^b -2 D_a x^a-\frac{W}{x^s-\bar{x}^s}, \ 
  \nabla_s\nabla_s W  =-2 \frac{\nabla_s W}{x^s-\bar{x}^s} ,  \\
 \nonumber \nabla_a\nabla_b W &= 2 C_{ab}(p^0 x^s-p^s)-\frac{2}{M}\left(C_{ab}-\frac{4 M_a M_b}{M}\right)W \\
\nonumber &-\frac{4}{M}\Big[M_a\big(p^0 C_{bc} x^c x^s-D_b x^s-C_{bc}x^c p^s\big)
+M_b\big(p^0 C_{ac} x^c x^s-D_a x^s-C_{ac}x^c p^s\big)\Big],\\
\nonumber \nabla_a\nabla_s W &= 2 \left( p^0 C_{ab} x^b-D_a\right)-\frac{2}{x^s-\bar{x}^s}\left(p^0 C_{ab} x^b x^s-D_a x^s-C_{ab} x^b p^s\right)\\
\nonumber & -\frac{2 M_a}{M}\left(p^0 C_{bc} x^b x^c-2D_b x^b\right)+\frac{2 M_a}{M}\frac{W}{x^s-\bar{x}^s} .
\end{align*}
where, for convenience, we have introduced the notations{\footnote{Our notations for $M_a,M,D_a$ and $D$
here are slightly different from the ones introduced in \cite{Tripathy:2005qp}.}}:
$D_a=C_{a b} p^b$ , $ D=C_{a b} p^a p^b$ and $C^{ab}C_{bc}=\delta^a_c$. To find the attractor point(s), 
we can now substitute the above expressions in the equations of motion:
\begin{eqnarray} \label{eom}
 g^{b\bar{c}} \nabla_a\nabla_b W \overline{\nabla_c W}+2 \nabla_a W \overline{W}+\partial_a g^{b\overline{c}}\nabla_b W\overline{\nabla_c W}+g^{s\bar{s}} \nabla_a\nabla_s W\overline{\nabla_sW} &=& 0 \ , \cr
 g^{b\bar{c}}\nabla_s\nabla_b W \overline{\nabla_c W}+g^{s\bar{s}}\nabla_s\nabla_s W \overline{\nabla_s W}+2 \nabla_s W \overline{W}+\partial_s g^{s \bar{s}}\nabla_s W \overline{\nabla_s W} &=& 0 \ . 
\end{eqnarray}
Substituting the expressions for the individual terms, the above equations become extremely complicated and it is 
hard to solve them in general without making any further assumption. Taking a clue from the supersymmetry 
preserving solutions we will first consider the simplest ansatz: $x^a = p^a t, x^s = p^s t$ (with $t= t_1 + i t_2$) to 
solve the equation of motion (\ref{eom}). The solution has been carried out in Ref.\cite{Tripathy:2005qp}. In the 
following we will briefly summarize the results.

We notice that the equations take a particularly simple form after a rescaling of the quantities $p^0,q_0$ 
and $t$ by $\tilde{p}^0 \sqrt{D p^s}, \nicefrac{\tilde{q}_0}{({\tilde{p}}^0)^2}$ and $\nicefrac{\tilde{t}}{\tilde{p}^0 \sqrt{D p^s} }$ respectively. The 
resulting equations depend only on a single parameter ($\tilde{q}_0$) and we find:
\begin{eqnarray} \label{oldeq}
A_2(B_1 - B_2) + B_2 (A_1 - A_2) = 0 \cr
A_2(A_1 + A_2) - B_2 (B_1 + B_2) = 0 
\end{eqnarray}
where for convenience, we have used the following notations:
\begin{eqnarray}
A_1 &=& (\tilde{q}_0 - 3 \tilde{t}_1^2 + \tilde{t}_1^3 + 3 \tilde{t}_2^2  - 3 \tilde{t}_1 \tilde{t}_2^2) \cr
A_2 &=& (\tilde{q}_0 - 3 \tilde{t}_1^2 + \tilde{t}_1^3 - \tilde{t}_2^2 + \tilde{t}_1 \tilde{t}_2^2) \cr
B_1 &=& \tilde{t}_2 ( 6 \tilde{t}_1 - 3 \tilde{t}_1^2 + \tilde{t}_2^2) \cr
B_2 &=& \tilde{t}_2 (\tilde{t}_2^2 - 2 \tilde{t}_1 + \tilde{t}_1^2)
\end{eqnarray}
The solution for which $\nabla_a W \neq 0$ is given by
\begin{eqnarray}\label{oldsoln}
\tilde{t}_1 &=& 1-\frac{u^{\nicefrac{1}{3}}+u}{1+u^{\nicefrac{4}{3}}}\\
\tilde{t}_2 &=& \frac{(u^2-1)}{u^{\nicefrac{1}{3}}(1+u^{\nicefrac{4}{3}})}
\end{eqnarray}
where $u$ is defined by the relation $\tilde{q}_0 u +  (u-1)^2 = 0$.{\footnote{Note that the attractor equations are invariant 
under the exchange $\tilde t_2 \rightarrow -\tilde t_2$. Since the imaginary part of the moduli must be negative at the attractor 
point, for a given set of $p^a$ the appropriate branch of solution for $\tilde t_2$ are chosen accordingly.}}

There is no particular reason for us to consider the above ansatz except for its simplicity.{\footnote{For example, this
ansatz plays no role in the case of non-supersymmetric attractors with vanishing central charge \cite{Bellucci:2007zi}.}
As we have seen in the previous section, for the two parameter model there are indeed solutions which are not of 
the above form. With arbitrary parameters it is in extremely difficult to construct the general solution without taking 
any specific ansatz.  However, in the present case, the scalar field $x^s$ plays a special role. Thus, it is natural to 
consider the ansatz $x^a = p^a t$ and $x^s = p^s j$.  The previous equations can be obtained by setting $j=t$. In 
the following we will proceed by treating $t$ and $j$ independent. To simplify our analysis we will again rescale 
the charges as well as the variables as in the previous case. To this end we set $p^0,q_0,t$ and $j$ as  
$ \tilde{p}^0 \sqrt{D p^s}, \nicefrac{\tilde{q}_0}{(\tilde{p}^0)^2}, \nicefrac{\tilde{t}}{\tilde{p}^0 \sqrt{D p^s} }$ 
and $\nicefrac{\tilde{j}}{\tilde{p}^0\sqrt{D p^s} }$  respectively. Once more we find that the equations are 
dependent only upon the charge $\tilde{q}_0$. After some simplification (as outlined in the appendix \S A) 
we find the equations of motion to take the form:
\begin{eqnarray}\label{neweqs}
 \nonumber X_2 (Y_3-Y_1)+Y_2(X_3-X_1) &=& 0,\\
 \nonumber X_2(X_1+X_3)-Y_2(Y_1+Y_3) &=& 0,\\
 \nonumber 2 X_2 Y_2-(X_1 Y_3+X_3 Y_1) &=& 0,\\
 X_2^2-Y_2^2+X_1 X_3-Y_1 Y_3 &=& 0,
\end{eqnarray}
where we have introduced the quantities $X_i,Y_i, (i=1,2,3)$ as follows:
\begin{eqnarray} \label{variables}
 {\tilde{p}_0}^2 X_1 &=& \tilde{q}_0-{\tilde{t}_2}^2 (\tilde{j}_1-1)+\tilde{t}_1\{\tilde{t}_1({\tilde{j}}_1-1)-2 \tilde{j}_1\}-2\tilde{t}_2\tilde{j}_2 (\tilde{t}_1-1), \nonumber\\
 {\tilde{p}_0}^2Y_1 &=& 2\tilde{t}_2\{\tilde{j}_1(\tilde{t}_1-1)-\tilde{t}_1\}+\tilde{j}_2 \{\tilde{t}_1(\tilde{t}_1-2)-{\tilde{t}_2}^2\}, \nonumber\\
 {\tilde{p}_0}^2 X_2 &=&  \tilde{q}_0+{\tilde{t}_2}^2(\tilde{j}_1-1)+\tilde{t}_1\{\tilde{t}_1(\tilde{j}_1-1)-2 \tilde{j}_1\}, \nonumber\\
 {\tilde{p}_0}^2 Y_2 &=&-\tilde{j}_2\{\ \tilde{t}_1(\tilde{t}_1-2)+{\tilde{t}_2}^2\},\nonumber\\
 {\tilde{p}_0}^2X_3 &=& \tilde{q}_0-{\tilde{t}_2}^2 (\tilde{j}_1-1)+\tilde{t}_1\{\tilde{t}_1(\tilde{j}_1-1)-2 \tilde{j}_1\}+2\tilde{t}_2\tilde{j}_2 (\tilde{t}_1-1), \nonumber\\
 {\tilde{p}_0}^2Y_3 &=& \tilde{j}_2\{\tilde{t}_1(\tilde{t}_1-2)-{\tilde{t}_2}^2\}-2\tilde{t}_2 \{\tilde{j}_1(\tilde{t}_1-1)-\tilde{t}_1\} .
\end{eqnarray}

As expected, we recover Eqs.(\ref{oldeq}) upon setting $\tilde{j}=\tilde{t}$. However, there is a possibility of obtaining 
genuinely new solutions from the above equations by treating $\tilde j$ and $\tilde t$ as independent variables. It is 
straightforward to solve Eq.(\ref{neweqs}) to find both supersymmetric as well non-supersymmetric 
solutions. The supersymmetric solution obtained from these equations are identical to the one obtained
from Eq.(\ref{oldeq}). The non-supersymmetric solution is given by
\begin{eqnarray}\label{newsoln}
 \tilde{t}_1 = \frac{\tilde{q}_0}{2}, \  \tilde{t}_2 = \frac{1}{2}\sqrt{\tilde q_0(\tilde{q}_0-4)},  \ 
\tilde{j}_1 = \frac{\tilde{q}_0(\tilde{q}_0-3)}{2 +\tilde{q}_0(\tilde{q}_0-4)}, \  \tilde{j}_2 = \frac{\sqrt{\tilde{q}_0(\tilde{q}_0-4)}}{(2 + \tilde{q}_0(\tilde{q}_0-4))}. 
\end{eqnarray}
In terms of the original variables before rescaling this solution takes the form:
\begin{eqnarray}
t &=& \frac{p^0 q_0}{2 D p^s}+ i \frac{\sqrt{q_0 ({p^0}^2 q_0-4 D p^s)}}{2 D p^s},\\
j&=& \frac{p^0 q_0 ({p^0}^2 q_0-3 D p^s)}{2 D^2 {p^s}^2-4 D p^s{p^0}^2 q_0+{p^0}^4{q_0}^2 }
+i \frac{D p^s \sqrt{q_0 ({p^0}^2 q_0-4 D p^s)}}{2 D^2 {p^s}^2-4 D p^s{p^0}^2 q_0+{p^0}^4{q_0}^2 }.
\end{eqnarray}
Note that the above solution as well as the one given in Eq.(\ref{oldsoln}) exist in the domain of the 
charge lattice for which $\left({p^0}^2 q_0^2 - 4 D p^s q_0\right) > 0$. In contrast, the supersymmetric 
attractors exists in the mutually exclusive domain $\left({p^0}^2 q_0^2 - 4 D p^s q_0\right) < 0$.

The entropy of the black hole can be computed from the formula $S = \pi V_0$ where $V_0$ is
the  value of the effective black hole potential at the attractor point. We find 
\begin{eqnarray}
S=\pi \sqrt{\left({p^0}^2 q_0^2 - 4 D p^s q_0\right)} \ , 
\end{eqnarray}
which, interestingly, is also identical to the entropy of the  non-supersymmetric attractor given in Eq.(\ref{oldsoln}).

Clearly this is a distinct solution than the one described in Eq.(\ref{oldsoln}) and hence we have multiple single
centered attractors (with non-vanishing axionic part) having the same charge configuration.  
%
%
One criteria which might distinguish the solutions 
form one another is stability. For the first solution (\ref{oldsoln}), there exist $(n+1)$ massive modes and the 
remaining $(n-1)$ fields become zero modes. For the solution (\ref{newsoln}) we need to compute the mass 
matrix and diagonalize it to find the stable directions.

The computation of the mass matrix has been discussed in detail in appendix \S B. Using the explicit 
expressions for the elements of the matrices $\Sigma_0$,$\Sigma_1$,$\Sigma_2$ and $\Sigma_3$, 
the mass matrix is found to have the form:
\small{\begin{flalign}
 &M = e^{K_0} \frac{  8 \left({p^0}^2q_0^2-4 D p^s q_0\right)}{\alpha D}	\begin{pmatrix}
											     \frac{4 D_a D_d}{D} & \frac{2 D_a}{p^s} & 0 & \beta \frac{D_a}{ps}\\
											     \frac{2 D_a}{p^s} & D \frac{\alpha^2}{4 {p^s}^2} & 0 & 0\\
											     0 & 0 & 2\big(2\frac{D_a D_d}{D}-C_{ad}\big) & 0\\
											      \beta \frac{D_a}{p^s} & 0& 0&D \frac{\alpha^2}{4 {p^s}^2}
											      \end{pmatrix},
\end{flalign}}
where we have used the notations: 
$$\alpha=\frac{2D^2{p^s}^2-4Dp^s{p^0}^2q_0+{p^0}^4q_0^2}{(Dp^s)^2}
\ {\rm  and } \  \beta=\frac{({p^0}^2q_0-2Dp^s)\sqrt{{({p^0}^4q_0^2-4Dp^sp^0}^2q_0)}}{(Dp^s)^2} \ . $$
Note that, for the non-supersymmetric solution, the numerator in the pre-factor of the mass matrix is positive (which also implies 
that $\alpha > 0)$. We now need to find the eigenvalues of the above matrix. 
By a trivial change of basis the above mass matrix 
can be brought into a block diagonal form of the type 
$$ 8\ e^{K_0}  \left({p^0}^2q_0^2-4 D p^s q_0\right)\frac{ 1}{\alpha D} 
\ \left(\begin{matrix} M_u & 0 \cr 0 & M_d \end{matrix}\right) $$
where the $(n-1)\times(n-1)$ matrix $(\nicefrac{1}{\alpha D}) M_u$ is given by a positive multiple of  the 
moduli space metric of $K3$ and hence it possesses $(n-1)$ positive eigenvalues for any smooth $K3$ 
surface. On the other hand, the $(n+1)\times(n+1)$ matrix $M_d$ is given by
\small{\begin{flalign}
 & M_d = \begin{pmatrix}
											     \frac{4 D_a D_d}{D} & \frac{2 D_a}{p^s} & \beta \frac{D_a}{ps}\\
											     \frac{2 D_a}{p^s} & D \frac{\alpha^2}{4 {p^s}^2}  & 0\\
											      \beta \frac{D_a}{p^s} & 0 &D \frac{\alpha^2}{4 {p^s}^2}
											      \end{pmatrix}.
\end{flalign}}

We will now obtain the eigenvalues of this  matrix. Consider first a vector of the form:
$$\begin{pmatrix} 0 \\  \varphi_1 \\  \varphi_2\end{pmatrix} . $$
This will be an eigenvector with eigenvalue $D (\nicefrac{\alpha}{2 p^s})^2$ provided $2 \varphi_1 + \beta \varphi_2 = 0 $. 
Vectors of the form
$$\begin{pmatrix} \  \omega_a \cr 0 \cr 0\end{pmatrix} $$ 
will be eigenfunctions with zero eigenvalues, provided $D_a \omega_a = 0$. Since $a= 1,\cdots,(n-1)$, there 
will be $(n-2)$ such  linearly independent vectors with zero eigenvalue. Finally, consider vectors of the form
$$\begin{pmatrix} \ \omega_a \cr \varphi_1 \cr \varphi_2 \end{pmatrix} \ . $$
Vectors of this type with $\varphi_1\neq0, \varphi_2\neq 0$ will satisfy the eigenvalue equation with eigenvalue $\lambda$
provided 
\begin{eqnarray}
\omega_a &=& D_a \cr
\lambda &=& \frac{4}{D} D_aD_a + \frac{1}{p^s} (2 \varphi_1 + \beta \varphi_2) \cr
\frac{2}{p^s} D_aD_a + D \left(\frac{\alpha}{2 p^s}\right)^2 \varphi_1 &=&
\frac{4 \varphi_1}{D} D_aD_a + \frac{\varphi_1}{p^s}(2 \varphi_1 + \beta \varphi_2) \cr
\frac{\beta}{p^s} D_aD_a + D \left(\frac{\alpha}{2 p^s}\right)^2 \varphi_2 &=&
\frac{4 \varphi_2}{D} D_aD_a + \frac{\varphi_2}{p^s}(2 \varphi_1 + \beta \varphi_2) 
\end{eqnarray}
The last two of the above equations are compatible with each other if and only if $\varphi_2 = (\nicefrac{\beta}{2}) \varphi_1$. 
Substituting for the above value of $\varphi_2$ we get a quadratic equation for $\varphi_1$ which admits the solutions 
$$ \varphi_{\mp} = - \frac{8 p^s}{D \alpha^2} D_aD_a , \frac{D}{2 p^s} \ , $$ 
with the respective eigenvalues 
$\lambda_{\mp} = 0 , \big((\nicefrac{4}{D}) D_aD_a + D(\nicefrac{\alpha}{2 p^s})^2\big)$. Thus the matrix 
$(\nicefrac{1}{\alpha D}) M_d$ has $(n-1)$ zero eigenvalues and two nonzero eigenvalues:
$(\nicefrac{\alpha}{(2p^s)^2})$ and 
$(\nicefrac{1}{\alpha})\big((\nicefrac{2}{D})^2 D_aD_a + (\nicefrac{\alpha}{2p^s})^2\big)$, both positive. 
To summarise, we observe that for the solution (\ref{newsoln}) the mass matrix admits $(n+1)$ positive 
eigenvalues and $(n-1)$ zero eigenvalues\cite{Nampuri:2007gv,Ferrara:2007tu,Ceresole:2007rq}. Thus, 
neither the entropy nor the number of zero modes distinguishes these two attractors from one another.

We would like to emphasise here that to have a well defined attractor solution, it is not sufficient to have 
a positive definite moduli space metric with positive definite mass matrix. The gauge kinetic terms also must 
be positive definite. The condition for this is expressed in terms of the real symplectic matrix:
\begin{eqnarray}
\mathcal{M}=\begin{pmatrix}
               Im \mathcal{N}+Re \mathcal{N}(Im \mathcal{N})^{-1}Re \mathcal{N} & -Re \mathcal{N}(Im \mathcal{N})^{-1}\\
               -(Im \mathcal{N})^{-1} Re \mathcal{N} & (Im \mathcal{N})^{-1}
              \end{pmatrix}
              \end{eqnarray}
where ${\mathcal N}$ is defined in terms of its matrix elements 
\begin{equation*}
 \mathcal{N}_{\Lambda \Sigma}=\bar{F}_{\Lambda \Sigma}
 +2 i \frac{(ImF_{\Lambda \Omega})(ImF_{\Pi\Sigma})X^\Omega X^\Pi}{(ImF_{\Omega \Pi})X^\Omega X^\Pi}.
\end{equation*}
The matrix ${\mathcal M}$ must be negative definite.

It is straightforward to compute the matrix ${\mathcal N}$ for the pre-potential (\ref{prepotx}). We find the real and 
imaginary parts to be 
 \begin{eqnarray*}
 Re \mathcal{N} =  \begin{pmatrix}
           2 C_{ab}x^a_r x^b_r x^s_r & -2 C_{db}x^b_r x^s_r & -C_{ab}x^a_rx^b_r\\
           -2 C_{cb}x^b_r x^s_r  & 2 C_{cd}x^s_r & 2C_{cb}x^b_r\\
           -C_{ab}x^a_rx^b_r & 2C_{db}x^b_r & 0
          \end{pmatrix}
          \end{eqnarray*}
          and 
\begin{eqnarray*}
Im \mathcal{N} = (C_{\tilde{c}\tilde{d}}x^{\tilde{c}}_ix^{\tilde{d}}_ix^s_i) \begin{pmatrix}
				  (1+4 g_{a\bar{b}}x^a_r x^b_r+4 g_{s\bar{s}}(x^s_r)^2) & -4 g_{d\bar{b}}x^b_r & -\frac{x^s_r}{{x^s_i}^2}\\
				   -4 g_{c\bar b}x^b_r  & 4 g_{c\bar{d}} & 0\\
				   -\frac{x^s_r}{{x^s_i}^2} & 0 & \frac{1}{{x^s_i}^2}
                                   \end{pmatrix}
                                   \end{eqnarray*}
respectively. Here we have used the notation $x^a=\frac{X^a}{X^0}=x^a_r+i x^a_i$ and $x^s=\frac{X^s}{X^0}=x^s_r+i x^s_i$.
Substituting the above and using the ansatz $x^a=p^a t=p^a(t_1+i t_2)$ and $x^s=p^s j=p^s(j_1+i j_2)$, we find the 
various elements of the matrix ${\mathcal M}$ to be given by
\begin{eqnarray*}
(Im\mathcal{N})^{-1}= \frac{1}{D p^s t_2^2 j_2}\begin{pmatrix}
           1 & p^b t_1 & p^s j_1\\
           p^a t_1 & p^a p^b t_1^2+\frac{1}{4}g^{a\bar{b}}& p^a p^s t_1 j_1\\
           p^s j_1 & p^b p^s t_1 j_1 & p_s^2 j_1^2+\frac{1}{4}g^{s\bar{s}}
          \end{pmatrix} \ , 
          \end{eqnarray*}
\begin{flalign}& \nonumber (Im\mathcal{N})^{-1}Re\mathcal{N}=(Re\mathcal{N}(Im\mathcal{N})^{-1}))^T\\\nonumber
&= \frac{1}{D p^s t_2^2 j_2}\begin{pmatrix}
           -D p^s t_1^2j_1 & 2 D_b p^s t_1 j_1 & D t_1^2\\
           -D p^a p^s t_1 j_1 (t_1^2+t_2^2) & 2  p^a D_b p^s j_1 (t_1^2+t_2^2)-D p^s t_2^2j_1\delta^a_b & D p^a t_1(t_1^2+t_2^2)\\
           -D {p^s}^2t_1^2(j_1^2+j_2^2) & 2 D_b {p^s}^2t_1(j_1^2+j_2^2) & D p^s t_1^2 j_1
           \end{pmatrix}\end{flalign}
           and
\begin{flalign} & \nonumber 
{{t_2}^2j_2} \big(Im \mathcal{N}+Re \mathcal{N}(Im \mathcal{N})^{-1}Re \mathcal{N}\big) = \\\nonumber
& \begin{pmatrix}
D p^s(t_1^2+{t_2}^2)^2(j_1^2+{j_2}^2) & -2 D_b p^s t_1(t_1^2+{t_2}^2)(j_1^2+{j_2}^2) & -D j_1 (t_1^2+{t_2}^2)^2\\
-2 D_a p^s t_1(t_1^2+{t_2}^2)(j_1^2+{j_2}^2) & 
(j_1^2+{j_2}^2)\Big(\frac{4 D_a D_b p^s}{D}(t_1^2+{t_2}^2)-2C_{ab}p^s {t_2}^2\Big) & 2 D_a t_1 j_1(t_1^2+{t_2}^2)\\
-D j_1 (t_1^2+{t_2}^2)^2 & 2 D_b t_1 j_1(t_1^2+{t_2}^2)& \frac{D(t_1^2+{t_2}^2)^2}{p^s}
          \end{pmatrix}\end{flalign}
          
For axion-free black holes, real part of the matrix ${\mathcal N}$ vanishes and the imaginary part is proportional to 
the moduli space metric with a negative proportionality factor. Thus the matrix ${\mathcal M}$ in this case is negative 
whenever the moduli space matrix is positive definite. Unfortunately, there is no simple way to diagonalise ${\mathcal M}$
in the present of axionic part. We have numerically computed the eigenvalues for a wide range of charges and with 
specific choice of the intersection matrix $C_{ab}$ and found ${\mathcal M}$ to be negative definite.

Before closing this session, we would like to point out that the appearance of multiple non-supersymmetric
attractors specific to our less specific ansatz $x^a = p^a t, x^s = p^s j$  seems to be a distinctive feature of the 
$D0-D4-D6$ configuration. Upon setting the $D6$ charge to  zero in Eqs.(\ref{oldsoln}) and (\ref{newsoln}) we 
find identical expression for the $D0-D4$ solution. On the other hand, while the solution (\ref{oldsoln}) gives a 
smooth $D0-D6$ solution \cite{Nampuri:2007gv,Dominic:2011st} in the limit $p^a,p^s \rightarrow 0$,  
Eqs.(\ref{newsoln}) becomes singular in this limit. Thus the solution (\ref{newsoln}) exists only for finite nonzero 
values of $D4$ charges.

\section{Conclusion}
In this paper we have studied multiple attractors in ${\cal N}=2$ supergravity obtained from the
compactification of type $IIA$ string theory on a Calabi-Yau manifold. Using the $4D-5D$ correspondence 
of black hole critical points, we constructed supersymmetric as well as non-supersymmetric attractors in
four dimensions. Further, by making some specific assumption on the intersection numbers we studied the 
attractor equations for spherically symmetric, extremal black holes with arbitrary number of vector multiplets. 
To simplify the analysis we assumed a simple ansatz for the scalar fields. Interestingly we found a unique 
supersymmetric attractor and  two distinct single centered non-supersymmetric attractors with the same 
charge configurations. These multiple non-supersymmetric solutions with the same charge configurations 
share many common properties. In particular, we found that the entropy for the corresponding black holes 
are the same and also they share the same number of zero modes.

For supersymmetric attractors in five dimensions the multiplicity arises because there exists several disjoint 
branches of moduli space \cite{Kallosh:1999mb}. Attractors in each of these branches are unique 
\cite{Wijnholt:1999vk}. These attractors give rise 
to both supersymmetric as well as non-supersymmetric axion free attractors in four dimensions upon 
dimensional reduction. Thus the multiplicity of these four dimensional attractors can be understood using
the $4D-5D$ correspondence. 

The multiple axionic $D0-D4-D6$ attractors  are less understood. First, there is neither any obvious symmetry 
relating both the solutions constructed here, nor are they related to the known $D0-D4-D6$ supersymmetric 
configuration by electromagnetic duality transformation. What is more puzzling is that  these multiple solutions appear to
exist even when the moduli space is connected. Clearly a more detailed  analysis is required to understand
these solutions fully. It would also be interesting to investigate the existence of multiple axionic attractors in the 
supersymmetric sector.

There are several other related issues that deserves future study. It would be interesting to explore the existence of 
multiple attractor points for a more general Calabi-Yau manifold without imposing any restriction on the 
intersection numbers and with a less restrictive ansatz for the scalar fields. A 
related issue is to understand the full set of non-supersymmetric attractor points for a given charge 
configuration in a particular Calabi-Yau compactification.  It would also be interesting to see what happens 
to the new solution we find when stringy corrections are included. Even for the simplest case of $D0-D4$
black holes the stringy correction introduces richer space of attractor solutions \cite{Bellucci:2007eh}. 
The stability conditions also change in an interesting way \cite{Bellucci:2008tx,Dominic:2011st}.  Adding 
$D6$ branes to this configuration will certainly enhance this already rich structure of non-supersymmetric 
attractors  as we now have a new ansatz to explore the solutions.  We hope to study some of these issues 
in near future. 

\section{Acknowledgment}
We would like to thank Samrat Bhowmick and Karthik Inbasekar for helpful discussions. One of us (PD) 
gratefully acknowledges the hospitality provided by the Department of Physics, IIT Madras in the initial 
stage of this project.

\appendix

\section{The Attractor Equation}
In this appendix, we will outline some of the computational details required to find the non-supersymmetric
attractor when we use the ansatz:
\begin{eqnarray}
x^a = p^a t ,  \ {\rm and} \ x^s = p^s j \ .
\end{eqnarray} 
We denote $t_1, j_1$ to be the real parts and $t_2,j_2$ to be the imaginary parts of $t,j$ respectively.
With this ansatz, the derivatives of the K\"ahler potential are given by:
\begin{equation}
\partial_a K = \frac{i D_a}{D t_2}, \ 
\partial_s K = \frac{i}{2 p^s j_2} \ . 
\end{equation}
The metric on the moduli space can be shown to have the form
\begin{eqnarray}
g_{a \bar{b}}&=&  \frac{1}{2 D {t_2}^2} \left(2 \frac{D_a D_b}{D}-C_{a b}\right) \ , \\
g_{s \bar{s}}&=&  \frac{1}{(2 p^s j_2)^2} \ , \ g_{a\bar s} = 0 \ , 
\end{eqnarray}
with its inverse 
\begin{eqnarray}
g^{a \bar{b}} &=& 2 {t_2}^2 D \left(\frac{2}{D}p^a p^b-C^{ab}\right)\\
g^{s \bar{s}} &=& (2 p^s j_2)^2 \ , g^{a\bar s} = 0 \ . 
\end{eqnarray}
In the above we have introduced the notation $D_a=C_{a b} p^b$ , $ D=C_{a b} p^a p^b$ and 
$C^{ab}C_{bc}=\delta^a_c$.
We also compute derivatives of the metric, which will be of use in evaluating the equations of motion:
\begin{eqnarray}
\partial_a g^{b \bar{c}} &=& 2 i t_2 (D_a C^{bc}-\delta_a^b p^c-\delta_a^c p^b)\\
\partial_s g^{s \bar{s}} &=& -4 i p^s j_2 .
\end{eqnarray}
Using the above mentioned ansatz, we can simplify the superpotential and it's covariant derivatives and express
them in terms of the rescaled variables. We find:
\begin{flalign}
 & W = X_1+i Y_1,\nonumber
 \\
& \nabla_a W=\frac{\tilde{p}^0 p^s}{\tilde{t}_2\sqrt{Dp^s}}D_a\left(Y_2+i X_2\right),\nonumber
\\
& \nabla_s W= \frac{\tilde{p}^0}{2\tilde{j}_2\sqrt{Dp^s}}D(Y_3+i X_3),\nonumber
\\
& \nabla_a \nabla_b W
=\frac{(\tilde{p}^0)^2 p^s}{2\tilde{t}_2^2}
\left[\left\{2\left(C_{ab}-2\frac{D_a D_b}{D}\right)X_2-C_{ab} X_3\right\}
-i \left\{2\left(C_{ab}-2\frac{D_a D_b}{D}\right)Y_2+C_{ab} Y_3\right\}\right],\nonumber
\\
& \nabla_a\nabla_s W = -\frac{(\tilde{p}^0)^2}{2\tilde{t}_2\tilde{j}_2}D_a(X_2+i Y_2),\nonumber
\\
&\nabla_s \nabla_s W=-\frac{(\tilde{p}^0)^2}{2 p^s \tilde{j}_2^2}D(X_3-i Y_3),
\end{flalign}
The expressions for $X_i$ and $Y_i,  (i=1,\cdots,3)$ are defined in Eq.(\ref{variables}). The individual terms in the equation of motion can now be computed as follows:
\begin{flalign}
 & g^{b\bar{c}}\nabla_a\nabla_bW\overline{\nabla_cW}=-\frac{p^0p^sD_a}{\tilde{t}_2\sqrt{Dp^s}}\left(Y_2-i X_2\right)\left[\left(2X_2+X_3\right)-i\left(2Y_2-Y_3\right)\right],\cr
 & 2\nabla_aW\overline{W}=2\frac{p^0p^sD_a}{\tilde{t}_2\sqrt{Dp^s}}\left(Y_2+iX_2\right)\left(X_1-iY_1\right),\cr
 & \partial_a g^{b\bar{c}}\nabla_b W\overline{\nabla_c W}=- 2 i \frac{p^0p^sD_a}{\tilde{t}_2\sqrt{Dp^s}}\left(X_2^2+Y_2^2\right),\cr
 & g^{s\bar{s}}\nabla_a\nabla_s W\overline{\nabla_s W} =-\frac{p^0p^sD_a}{\tilde{t}_2\sqrt{Dp^s}}\left(X_2+i Y_2\right)\left(Y_3-iX_3\right);\nonumber
\end{flalign}
and
\begin{flalign}
  & g^{b\bar{c}}\nabla_s\nabla_b W\overline{\nabla_c W}=i\frac{Dp^0}{\tilde{j}_2\sqrt{Dp^s}}\left(X_2+iY_2\right)^2,\cr
 & g^{s\bar{s}}\nabla_s\nabla_s W\overline{\nabla_sW}=i \frac{Dp^0}{\tilde{j}_2\sqrt{Dp^s}}\left(X_3^2+Y_3^2\right),\cr
  & 2\nabla_sW\overline{W}=i \frac{Dp^0}{\tilde{j}_2\sqrt{Dp^s}}\left(X_3-iY_3\right)\left(X_1-iY_1\right),\cr
  & \partial_s g^{s\bar{s}}\nabla_sW\overline{\nabla_sW}=-i \frac{Dp^0}{\tilde{j}_2\sqrt{Dp^s}}\left(X_3^2+Y_3^2\right).
 \end{flalign}
Adding the above terms and simplifying we find the equations of motion as given in (\ref{neweqs}).

\section{The Mass Matrix}

In this appendix we will evaluate the mass matrix for our attractor solution. Expanding the effective black hole 
potential around the attractor point, we find the quadratic terms to be of the form:
\small{\begin{flalign}
 & 2\partial_a \partial_{\bar{d}} V \left(y^{1a}y^{1d}+y^{2a}y^{2d}\right)+4 Re(\partial_a \partial_{\bar{s}}V) \left(y^{1a}y^{1s}+y^{2a}y^{2s}\right)
 -4Im(\partial_a \partial_{\bar{s}}V) \left(y^{2a}y^{1s}-y^{1a}y^{2s}\right)
 \nonumber \\ &+ 2\partial_s \partial_{\bar{s}}V \left((y^{1s})^2+(y^{2s})^2\right)
 +2 Re(\partial_a\partial_d V)\left(y^{1a}y^{1d}-y^{2a}y^{2d}\right)-2 Im(\partial_a\partial_d V)\left(y^{2a}y^{1d}+y^{1a}y^{2d}\right)
 \nonumber\\& +4 Re(\partial_a\partial_s V)\left(y^{1a}y^{1s}-y^{2a}y^{2s}\right)-4 Im(\partial_a\partial_s V)\left(y^{2a}y^{1s}+y^{1a}y^{2s}\right),
\end{flalign}}
where, we set $x^a=x_0^a+y^{1a}+i y^{2a}$ and $x^s=x_0^s+y^{1s}+i y^{2s}$. From the above we notice that 
the mass matrix can be recast as 
\begin{equation}
 M=I \otimes \Sigma_0 - \sigma_1\otimes  \Sigma_1+i \sigma_2 \otimes \Sigma_2  + \sigma_3 \otimes \Sigma_3,
\end{equation}
where, $\sigma_1$, $\sigma_2$ and $\sigma_3$ are the Pauli matrices and the $n\times n$ matrices $\Sigma_i$
 $(i = 0,\cdots,3)$ are given by
\small{\begin{align*}
 \Sigma_0 &=   \begin{pmatrix}
            2\partial_a\partial_{\bar{d}} V & 2 Re(\partial_a \partial_{\bar{s}} V)\\
            2 Re(\partial_a \partial_{\bar{s}} V) & 2\partial_s \partial_{\bar{s} }V
            \end{pmatrix}, \ 
& \Sigma_1 &= \begin{pmatrix}
            2Im(\partial_a\partial_d V) & 2 Im(\partial_a \partial_s V)\\
            2 Im(\partial_a \partial_s V) & 0
            \end{pmatrix},\\
\Sigma_2 & =\begin{pmatrix}
            0 & 2 Im(\partial_a \partial_{\bar{s}} V)\\
            -2 Im(\partial_a \partial_{\bar{s}} V) & 0
            \end{pmatrix}, \ 
& \Sigma_3 &= \begin{pmatrix}
            2Re(\partial_a\partial_d V )& 2 Re(\partial_a \partial_s V)\\
            2 Re(\partial_a \partial_s V) & 0
            \end{pmatrix}.
  \end{align*}}
In order to obtain the mass matrix, we need to evaluate each of these terms at the attractor point. In the following
we will outline some of the intermediate steps in evaluating the mass matrix. A straightforward
differentiation gives the second derivative terms of the effective black hole potential as \cite{Tripathy:2005qp}:
\small{
\begin{flalign}
  e^{-K_0}\partial_a\partial_d V &= g^{b\bar{c}}\nabla_a \nabla_b \nabla_d W\overline{\nabla_c W}+
 g^{s\bar{s}}\nabla_a\nabla_s\nabla_d W\overline{\nabla_s W}+\partial_a g^{b\bar{c}} \nabla_b \nabla_d W \overline{\nabla_c W} 
\nonumber \\ & +\partial_d g^{b\bar{c}} \nabla_b \nabla_a W \overline{\nabla_c W}+3 \nabla_a \nabla_d W \overline{W}
 +\partial_a \partial_d g^{b\bar{c}} \nabla_b W \overline{\nabla_c W}-g^{b\bar{c}}\partial_a g_{d\bar{c}}\nabla_b W \overline{W},\nonumber
 \\
  e^{-K_0}\partial_a\partial_s V &=  g^{b\bar{c}}\nabla_a \nabla_b \nabla_s W\overline{\nabla_c W}
 +g^{s\bar{s}}\nabla_a\nabla_s\nabla_s W\overline{\nabla_s W}
 +\partial_a g^{b\bar{c}} \nabla_b \nabla_s W\overline{\nabla_c W} 
 \nonumber\\ &+\partial_s g^{s\bar{s}} \nabla_s \nabla_a W\overline{\nabla_s W} +3 \nabla_a \nabla_s \overline{W}, \nonumber
 \\
  e^{-K_0}\partial_a\partial_{\bar{d}} V &= g^{b\bar{c}} \nabla_a \nabla_b W\overline{\nabla_c\nabla_d W}
 +g^{s\bar{s}}\nabla_a\nabla_s W \overline{\nabla_s \nabla_d W} +2|W|^2 g_{a\bar{d}}\nonumber
 \\ &+g^{b\bar{c}}\nabla_b W\overline{\nabla_c W} g_{a\bar{d}} + g^{s\bar{s}} \nabla_s W \overline{\nabla_s W} g_{a\bar{d}}+\partial_a g^{b\bar{c}} \nabla_b W \overline{\nabla_c \nabla_d W}\nonumber
  \\ &+\partial_{\bar{d}} g^{b\bar{c}} \nabla_a \nabla_b W \overline{\nabla_c W}+
 3 \nabla_a W \nabla_d W+\nabla_a \nabla_{\bar{d}} g^{b\bar{c}}\nabla_b W \overline{\nabla_c W},\nonumber
 \\
 e^{-K_0} \partial_a \partial_{\bar{s}} V &= g^{b\bar{c}}\nabla_a\nabla_b W\overline{\nabla_c\nabla_s W}
 +g^{s\bar{s}}\nabla_a\nabla_s W\overline{\nabla_s \nabla_s W}+\partial_a g^{b\bar{c}} \nabla_b W\overline{\nabla_c \nabla_s W}\nonumber\\
 & +\partial_{\bar{s}}g^{s\bar{s}}\nabla_a\nabla_s W\overline{\nabla_s  W}+3 \nabla_a W\overline{\nabla_s W},\nonumber
 \\
  e^{-K_0}\partial_s\partial_{\bar{s}} V &= g^{b\bar{c}} \nabla_s \nabla_b W\overline{\nabla_c \nabla_s W}
 +g^{s\bar{s}}\nabla_s\nabla_s W\overline{\nabla_s \nabla_s W}+2 |W|^2 g_{s\bar{s}}+g^{b\bar{c}} \nabla_b W\overline{\nabla_c W} g_{s\bar{s}}\nonumber
\\&+\partial_s g^{s\bar{s}} \nabla_s W\overline{\nabla_s \nabla_s W}+\partial_{\bar{s}} g^{s\bar{s}}\nabla_s\nabla_s W\overline{\nabla_s W}
 +4 \nabla_s W\overline{\nabla_s W}+\partial_s\partial_{\bar{s}} g^{s\bar{s}}\nabla_s W\overline{\nabla_s W},
 \nonumber
 \\
 e^{-K_0}\partial_s\partial_s V &=g^{b\bar{c}}\nabla_s\nabla_b\nabla_s W\overline{\nabla_c W}+
 g^{s\bar{s}}\nabla_s \nabla_s \nabla_s W\overline{\nabla_s W}+2 \partial_s g^{s\bar{s}} \nabla_s \nabla_s W \overline{\nabla_s W}
 \nonumber
 \\ & + 3 \nabla_s\nabla_s W \overline{W}+\partial^2_s g^{s\bar{s}}\nabla_s W\overline{\nabla_s W}
 -g^{s\bar{s}}\partial_s g_{s\bar{s}} \nabla_s W\overline{W},
 \end{flalign}}
 where $K_0$ is the value of the K\"{a}hler potential at the attractor point.
 
 We will now evaluate each of the above expressions separately. The individual terms in 
 $e^{-K_0}\partial_a \partial_d V$ are given by
 \small{\begin{flalign}
 & g^{b\bar{c}}\nabla_a \nabla_b \nabla_d W\overline{\nabla_c W} = 2\frac{(\tilde{p}^0)^2p^s}{\tilde{t}_2^2}(Y_2-i X_2)
 \Big[\left\{\frac{1}{2}C_{ad} ( Y_3+3 Y_2 )+\frac{D_a D_d}{D}(Y_3-3 Y_2)\right\}\nonumber
  \\&-i\left\{\frac{1}{2}C_{ad}(X_3-3X_2)+\frac{D_a D_d}{D}(X_3+3 X_2)\right\}\Big]\nonumber,
 \\
 & g^{s\bar{s}}\nabla_a\nabla_s\nabla_d W\overline{\nabla_s W} = -\frac{1}{2}\frac{(\tilde{p}^0)^2 p^s}{\tilde{t}_2^2}(Y_3-i X_3)
  \Big[\left\{2\left(C_{ad}-2\frac{D_a D_d}{D}\right) Y_2+C_{ad} Y_1\right\}\nonumber
  \\&-i \left\{2\left(C_{ad}-2\frac{D_a D_d}{D}\right) X_2-C_{ad} X_1\right\}\Big],\nonumber
 \\
& \partial_a g^{b\bar{c}} \nabla_b \nabla_d W \overline{\nabla_c W} = \partial_d g^{b\bar{c}} \nabla_b \nabla_a W \overline{\nabla_c W}
 =-\frac{(\tilde{p}^0)^2p^s}{\tilde{t}_2^2}(X_2+i Y_2)\nonumber
 \\& \Big[\left\{2\left(C_{ad}-2\frac{D_a D_d}{D}\right)X_2-C_{ad} X_3\right\} 
   -i\left\{2\left(C_{ad}-2\frac{D_a D_d}{D}\right)Y_2+C_{ad} Y_3\right\}\Big],\nonumber
 \\
 & 3 \nabla_a \nabla_d W \overline{W} = \frac{3}{2}\frac{(\tilde{p}^0)^2 p^s}{\tilde{t}_2^2}(X_1-i Y_1)\Big[\left\{2\left(C_{ad}-2\frac{D_a D_d}{D}\right)X_2-C_{ad} X_3\right\}\nonumber
  \\& -i\left\{2\left(C_{ad}-2\frac{D_a D_d}{D}\right)Y_2+C_{ad} Y_3\right\}\Big],\nonumber \\
& \partial_a \partial_d g^{b\bar{c}} \nabla_b W \overline{\nabla_c W} =\frac{(\tilde{p}^0)^2 p^s}{\tilde{t}_2^2}  \left(C_{ad}-2\frac{D_a D_d}{D}\right) (X_2^2+Y_2^2),\nonumber
 \\
 &-g^{b\bar{c}}\partial_a g_{d\bar{c}}\nabla_b W \overline{W} = i \frac{(\tilde{p}^0)^2 p^s}{\tilde{t}_2^2}  \left(C_{ad}-2\frac{D_a D_d}{D}\right)\left(Y_2+i X_2\right)
 \left(X_1-i Y_1\right).
\end{flalign}}
 
Adding  all these terms, we find a simple expression for $e^{-K_0}\partial_a\partial_d V$ which is listed towards the
end of this section.
Similarly, we will now evaluate the remaining terms in the second derivatives of the potential. Individual terms
in $e^{-K_0}\partial_a\partial_s V$ are evaluated to be
\small{\begin{flalign}
 & g^{b\bar{c}}\nabla_a \nabla_b \nabla_s W\overline{\nabla_c W}\nonumber\\
 & = -\frac{(\tilde{p}^0)^2}{2\tilde{t}_2\tilde{j}_2}D_a\left[\{2\left(X_2^2-Y_2^2\right)+X_1 X_2
 +Y_1 Y_2\}  +i\{4 X_2 Y_2+X_1 Y_2-X_2 Y_1\}\right],\nonumber
 \\
 & g^{s\bar{s}}\nabla_a\nabla_s\nabla_s W\overline{\nabla_s W} = -\frac{(\tilde{p}^0)^2}{\tilde{t}_2\tilde{j}_2}D_a\left[X_2 X_3-Y_2 Y_3
 + i\left(X_2 Y_3+X_3 Y_2\right)\right],\nonumber
 \\
& \partial_a g^{b\bar{c}} \nabla_b \nabla_s W\overline{\nabla_c W} = \frac{(\tilde{p}^0)^2}{\tilde{t}_2\tilde{j}_2} D_a\left( X_2^2-Y_2^2+ 2 i X_2 Y_2\right),\nonumber
 \\
& \partial_s g^{s\bar{s}} \nabla_s \nabla_a W\overline{\nabla_s W} = \frac{(\tilde{p}^0)^2}{\tilde{t}_2\tilde{j}_2} D_a\left[\left(X_2 X_3-Y_2 Y_3\right)
 +i\left(X_2 Y_3+X_3 Y_2\right)\right],\nonumber
 \\
 & 3 \nabla_a \nabla_s \overline{W} = -\frac{3}{2}\frac{(\tilde{p}^0)^2}{\tilde{t}_2\tilde{j}_2}D_a\left[\left(X_1 X_2+Y_1 Y_2\right)+i\left(X_1 Y_2-X_2 Y_1\right)\right].
 \end{flalign}}
Addition of all these terms gives the value of $e^{-K_0}\partial_a\partial_s V$.
Individual terms  in $e^{-K_0}\partial_s^2  V$ are given by
\small{\begin{flalign}
 g^{b\bar{c}}\nabla_s\nabla_b\nabla_s W\overline{\nabla_c W} &= \frac{(\tilde{p}^0)^2 D}{p^s \tilde{j}_2^2}\left(Y_2^2-X_2^2
 - 2 i X_2 Y_2\right),\nonumber
 \\
 g^{s\bar{s}}\nabla_s \nabla_s \nabla_s W\overline{\nabla_s W} &= -\frac{3}{2}\frac{(\tilde{p}^0)^2 D}{p^s \tilde{j}_2^2} \left(X_3^2+Y_3^2\right),\nonumber
 \\
 2 \partial_s g^{s\bar{s}} \nabla_s \nabla_s W \overline{\nabla_s W} &= 2 \frac{(\tilde{p}^0)^2 D}{p^s \tilde{j}_2^2}\left(X_3^2+Y_3^2\right),\nonumber
 \\
  3 \nabla_s\nabla_s W \overline{W} &= -\frac{3}{2}\frac{(\tilde{p}^0)^2 D}{p^s \tilde{j}_2^2}\left[\left(X_1 X_3-Y_1 Y_3\right)-i\left(X_1 Y_3+X_3 Y_1\right)\right],\nonumber
  \\
 \partial^2_s g^{s\bar{s}}\nabla_s W\overline{\nabla_s W} &=  -\frac{(\tilde{p}^0)^2 D}{2 p^s \tilde{j}_2^2}\left(X_3^2+Y_3^2\right),\nonumber
 \\
 -g^{s\bar{s}}\partial_s g_{s\bar{s}} \nabla_s W\overline{W} &= \frac{(\tilde{p}^0)^2 D}{2 p^s \tilde{j}_2^2}\left[\left(X_1 X_3-Y_1Y_3\right)
 -i\left(X_1 Y_3+X_3 Y_1\right)\right].
\end{flalign}}
From the above we find that $ e^{-K_0}\partial_s^2 V$ vanished upon using the equations of motion.
$e^{-K_0}\partial_a\partial_{\bar{d}} V$ contains the following terms
\small{\begin{flalign}
 & g^{b\bar{c}} \nabla_a \nabla_b W\overline{\nabla_c\nabla_d W}\nonumber
 \\ &= \frac{(\tilde{p}^0)^2 p^s}{2 \tilde{t}_2^2} \left[- \left(C_{ad}-2 \frac{D_a D_d}{D}\right)
 \left(4 X_2^2+4 Y_2^2+X_3^2+Y_3^2\right) +4 C_{ad} \left(X_2 X_3 -Y_2 Y_3\right)\right],\nonumber
 \\
 & g^{s\bar{s}}\nabla_a\nabla_s W \overline{\nabla_s \nabla_d W} = \frac{(\tilde{p}^0)^2 p^s}{ \tilde{t}_2^2}\frac{D_a D_d}{D} \left(X_2^2+Y_2^2\right),\nonumber
 \\
 & 2|W|^2 g_{a\bar{d}} = - \frac{(\tilde{p}^0)^2 p^s}{ \tilde{t}_2^2}\left(C_{ad}-2 \frac{D_a D_d}{D}\right)\left(X_1^2+Y_1^2\right),\nonumber
 \\
 & g^{b\bar{c}}\nabla_b W\overline{\nabla_c W} g_{a\bar{d}} = -\frac{(\tilde{p}^0)^2 p^s}{ \tilde{t}_2^2}\left(C_{ad}-2 \frac{D_a D_d}{D}\right)\left(X_2^2+Y_2^2\right),\nonumber
 \\
 & g^{s\bar{s}} \nabla_s W \overline{\nabla_s W} g_{a\bar{d}} = -\frac{(\tilde{p}^0)^2 p^s}{2\tilde{t}_2^2}\left(C_{ad}-2 \frac{D_a D_d}{D}\right)\left(X_3^2+Y_3^2\right),\nonumber
 \\
 & \partial_a g^{b\bar{c}} \nabla_b W \overline{\nabla_c \nabla_d W} = \frac{(\tilde{p}^0)^2 p^s}{\tilde{t}_2^2}
  \Bigg[\left\{2 \left(C_{ad}-2 \frac{D_a D_d}{D}\right)\left(X_2^2+Y_2^2\right)-C_{ad}\left(X_2 X_3-Y_2 Y_3\right)\right\}\nonumber
\\&  +i C_{ad}\left(X_2 Y_3+X_3 Y_2\right)\Big],\nonumber
  \\
  &\partial_{\bar{d}} g^{b\bar{c}} \nabla_a \nabla_b W \overline{\nabla_c W} 
  = \frac{(\tilde{p}^0)^2 p^s}{\tilde{t}_2^2} \Bigg[\left\{2 \left(C_{ad}-2 \frac{D_a D_d}{D}\right)\left(X_2^2+Y_2^2\right) -C_{ad}\left(X_2 X_3-Y_2 Y_3\right)\right\}\nonumber
  \\& -i C_{ad}\left(X_2 Y_3+X_3 Y_2\right)\Big],\nonumber
  \\
 & 3 \nabla_a W \nabla_d W = 3\frac{(\tilde{p}^0)^2 p^s}{\tilde{t}_2^2}\frac{D_a D_d}{D}\left(X_2^2+Y_2^2\right),\nonumber
  \\
 & \nabla_a \nabla_{\bar{d}} g^{b\bar{c}}\nabla_b W \overline{\nabla_c W} = -\frac{(\tilde{p}^0)^2 p^s}{\tilde{t}_2^2}
  \left(C_{ad}-2 \frac{D_a D_d}{D}\right)\left(X_2^2+Y_2^2\right).
\end{flalign}}
  Summing up the above terms and using equations of motion, we obtain a simple expression for 
  $e^{-K_0}\partial_a\partial_{\bar{d}} V$. Similarly,
the second derivative $ e^{-K_0} \partial_a \partial_{\bar{s}} V $ has the following terms
\small{\begin{flalign}
& g^{b\bar{c}}\nabla_a\nabla_b W\overline{\nabla_c\nabla_s W}\nonumber\\ &= \frac{(\tilde{p}^0)^2}{2\tilde{t}_2\tilde{j}_2}D_a\left[\{2\left(X_2^2-Y_2^2\right)+X_2 X_3+Y_2 Y_3\}
 -i\left(4 X_2 Y_2-X_2 Y_3+X_3 Y_2\right)\right],\nonumber
 \\
 & g^{s\bar{s}}\nabla_a\nabla_s W\overline{\nabla_s \nabla_s W} = \frac{(\tilde{p}^0)^2}{\tilde{t}_2\tilde{j}_2}D_a\left[\left(X_2 X_3-Y_2 Y_3\right)
 +i\left(X_2 Y_3+X_3 Y_2\right)\right],\nonumber
 \\
 & \partial_a g^{b\bar{c}} \nabla_b W\overline{\nabla_c \nabla_s W} = \frac{(\tilde{p}^0)^2}{\tilde{t}_2\tilde{j}_2}D_a\left(Y_2^2-X_2^2+ 2 i  X_2 Y_2\right),\nonumber
 \\
& \partial_{\bar{s}}g^{s\bar{s}}\nabla_a\nabla_s W\overline{\nabla_s  W} = -\frac{(\tilde{p}^0)^2}{\tilde{t}_2\tilde{j}_2}D_a \left[\left(X_2 X_3-Y_2 Y_3\right)
 +i\left(X_2 Y_3+X_3 Y_2\right)\right],\nonumber
 \\
& 3 \nabla_a W\overline{\nabla_s W} = \frac{3}{2}\frac{(\tilde{p}^0)^2}{\tilde{t}_2\tilde{j}_2}D_a\left[\left(X_2 X_3+ Y_2 Y_3\right)+i\left(X_2 Y_3-X_3 Y_2\right)\right].
\end{flalign}}
Finally, we evaluate terms in $e^{-K_0}\partial_s\partial_{\bar{s}} V$:
\small{\begin{flalign}
 & g^{b\bar{c}} \nabla_s \nabla_b W\overline{\nabla_c \nabla_s W} = \frac{(\tilde{p}^0)^2}{2 \tilde{j}_2^2 p^s}D\left(X_2^2+Y_2^2\right),\nonumber
 \\
 & g^{s\bar{s}}\nabla_s\nabla_s W\overline{\nabla_s \nabla_s W} = \frac{(\tilde{p}^0)^2}{ \tilde{j}_2^2 p^s}D\left(X_3^2+Y_3^2\right),\nonumber
 \\
& 2 |W|^2 g_{s\bar{s}} = \frac{(\tilde{p}^0)^2}{2 \tilde{j}_2^2 p^s}D\left(X_1^2+Y_1^2\right),\nonumber
 \\
& g^{b\bar{c}} \nabla_b W\overline{\nabla_c W} g_{s\bar{s}} = \frac{(\tilde{p}^0)^2}{2 \tilde{j}_2^2 p^s}D\left(X_2^2+Y_2^2\right),\nonumber
 \\
 & \partial_s g^{s\bar{s}} \nabla_s W\overline{\nabla_s \nabla_s W} = -\frac{(\tilde{p}^0)^2}{ \tilde{j}_2^2 p^s}D\left(X_3^2+Y_3^2\right),\nonumber
 \\
 & \partial_{\bar{s}} g^{s\bar{s}}\nabla_s\nabla_s W\overline{\nabla_s W} = -\frac{(\tilde{p}^0)^2}{ \tilde{j}_2^2 p^s}D\left(X_3^2+Y_3^2\right),\nonumber
 \\
& 4 \nabla_s W\overline{\nabla_s W} = \frac{(\tilde{p}^0)^2}{ \tilde{j}_2^2 p^s}D\left(X_3^2+Y_3^2\right),\nonumber
 \\
& \partial_s\partial_{\bar{s}} g^{s\bar{s}}\nabla_s W\overline{\nabla_s W} = \frac{(\tilde{p}^0)^2}{2\tilde{j}_2^2 p^s}D\left(X_3^2+Y_3^2\right).
\end{flalign}}
Adding  all these terms and using equations of motion, we get an expression for $ e^{-K_0}\partial_s\partial_{\bar{s}} V$.

To summarize,  the various terms in the mass matrix are found to be 
\begin{eqnarray}
 e^{-K_0}\partial_a\partial_d V &=& -2 \frac{(\tilde{p}^0)^2 p^s}{\tilde{t}_2^2} C_{ad}\left[\left(X_1 X_3+Y_1 Y_3\right)-i\left(X_3 Y_1-X_1 Y_3\right)\right]. \cr
  e^{-K_0}\partial_a\partial_s V &=& -2 \frac{(\tilde{p}^0)^2}{\tilde{t}_2\tilde{j}_2} D_a \left[\left(X_1 X_2+Y_1 Y_2\right)+i\left(X_1 Y_2-X_2 Y_1\right)\right]. \cr
e^{-K_0}\partial_s\partial_s V &=& 0 \cr
 e^{-K_0}\partial_a\partial_{\bar{d}} V &=& -2 \frac{(\tilde{p}^0)^2 p^s}{\tilde{t}_2^2}\left(C_{ad}-4 \frac{D_a D_d}{D}\right)\left(X_1^2+Y_1^2\right). \cr
  e^{-K_0} \partial_a \partial_{\bar{s}} V &= & 2 \frac{(\tilde{p}^0)^2}{\tilde{t}_2\tilde{j}_2}D_a\left[\left(X_2 X_3+Y_2 Y_3\right)+i\left(X_2 Y_3-X_3 Y_2\right)\right]. \cr
 e^{-K_0}\partial_s\partial_{\bar{s}} V &=& 2 \frac{(\tilde{p}^0)^2}{\tilde{j}_2^2 p^s} D \left(X_1^2+Y_1^2\right).
\end{eqnarray}
Note that, for the new solution (\ref{newsoln}), for which we are interested to find the mass matrix, we have 
$X_1 = - X_2 = - X_3$ and $Y_1 = Y_2 = - Y_3$. Using this we simplify the above equations to find:
\begin{eqnarray}
 e^{-K_0}\partial_a\partial_d V &=& 2 \frac{(\tilde{p}^0)^2 p^s}{\tilde{t}_2^2} C_{ad} \left(X_1^2 +Y_1^2 \right) \ ,  \cr
  e^{-K_0}\partial_a\partial_s V &=& 2 \frac{(\tilde{p}^0)^2}{\tilde{t}_2\tilde{j}_2} D_a \left[\left( X_1^2 - Y_1^2\right)
  - 2 i X_1  Y_1\right]. \cr
e^{-K_0}\partial_s\partial_s V &=& 0 \cr
 e^{-K_0}\partial_a\partial_{\bar{d}} V &=& -2 \frac{(\tilde{p}^0)^2 p^s}{\tilde{t}_2^2}\left(C_{ad}-4 \frac{D_a D_d}{D}\right)\left(X_1^2+Y_1^2\right). \cr
  e^{-K_0} \partial_a \partial_{\bar{s}} V &= & 2 \frac{(\tilde{p}^0)^2}{\tilde{t}_2\tilde{j}_2}D_a\left[\left(X_1^2 - Y_1^2 \right)+2 i X_1 Y_1 \right]. \cr
 e^{-K_0}\partial_s\partial_{\bar{s}} V &=& 2 \frac{(\tilde{p}^0)^2}{\tilde{j}_2^2 p^s} D \left(X_1^2+Y_1^2\right).
\end{eqnarray}
Using the solution (\ref{newsoln}), finally we get
\begin{flalign}
 & e^{-K_0}\partial_a\partial_d V=4 \tilde{q}_0 \frac{  p_s}{(\tilde{p}^0)^2}\frac{(\tilde{q}_0-4)}{(2+\tilde{q}_0(\tilde{q}_0-4))}C_{ad}\nonumber
 \\
 & e^{-K_0}\partial_a\partial_s V = \frac{4 \tilde{q}_0}{(\tilde{p}^0)^2}\frac{(\tilde{q}_0-4)}{(2+\tilde{q}_0(\tilde{q}_0-4))}
 \left(1-\frac{i}{2}(\tilde{q}_0-2)\sqrt{\tilde{q}_0(\tilde{q}_0-4)}\right)D_a\nonumber
 \\
 & e^{-K_0}\partial_s\partial_s V =0\nonumber
 \\
 & e^{-K_0}\partial_a\partial_{\bar{d}} V =4 \tilde{q}_0 \frac{p_s}{(\tilde{p}^0)^2}\frac{(\tilde{q}_0-4)}{(2+\tilde{q}_0(\tilde{q}_0-4))}\left(4\frac{D_a D_d}{D}-C_{ad}\right)\nonumber
 \\
 & e^{-K_0}\partial_a\partial_{\bar{s}} V =\frac{4 \tilde{q}_0}{(\tilde{p}^0)^2}\frac{(\tilde{q}_0-4)}{(2+\tilde{q}_0(\tilde{q}_0-4))}
 \left(1+\frac{i}{2}(\tilde{q}_0-2)\sqrt{\tilde{q}_0(\tilde{q}_0-4)}\right)D_a\nonumber
 \\
 & e^{-K_0}\partial_s\partial_{\bar{s}} V =\frac{\tilde{q}_0}{(\tilde{p}^0)^2p^s}(\tilde{q}_0-4)(2+\tilde{q}_0(\tilde{q}_0-4))D
\end{flalign}

\end{document}